\documentclass{emulateapj}

\begin{document}

\title{GALEX UV Color-Magnitude Relations and Evidence for Recent
Star Formation in Early-type Galaxies}

\author{
S. K. Yi,\altaffilmark{1,2,3}
S.-J. Yoon,\altaffilmark{2,3}
S. Kaviraj,\altaffilmark{3}
J.-M. Deharveng,\altaffilmark{4}
R. M. Rich,\altaffilmark{5}
S. Salim,\altaffilmark{5}
A. Boselli,\altaffilmark{4}
Y.-W. Lee,\altaffilmark{2} 
C. H. Ree,\altaffilmark{2}
Y.-J. Sohn,\altaffilmark{2} 
S.-C. Rey,\altaffilmark{2,6}
J.-W. Lee,\altaffilmark{11} 
J. Rhee,\altaffilmark{2,6}
L. Bianchi,\altaffilmark{7}
Y.-I. Byun,\altaffilmark{2}
J. Donas,\altaffilmark{4}
P. G. Friedman,\altaffilmark{6}
T. M. Heckman,\altaffilmark{7}
P. Jelinsky,\altaffilmark{12}
%Y.-W. Lee,\altaffilmark{1}
B. F. Madore,\altaffilmark{8,9}
R. Malina,\altaffilmark{4}
D. C. Martin,\altaffilmark{6}
B. Milliard,\altaffilmark{4}
P. Morrissey,\altaffilmark{6}
S. Neff,\altaffilmark{13}
%R. M. Rich,\altaffilmark{1}
D. Schiminovich,\altaffilmark{10}
O. Siegmund,\altaffilmark{12}
T. Small,\altaffilmark{6}
A. S. Szalay,\altaffilmark{7}
M. J. Jee,\altaffilmark{7}
S.-W. Kim,\altaffilmark{2}
T. Barlow,\altaffilmark{6}
K. Forster,\altaffilmark{6}
B. Welsh,\altaffilmark{12}
\& T. K. Wyder\altaffilmark{6}
}
\altaffiltext{1}{Email: yi@astro.ox.ac.uk}
\altaffiltext{2}{Center for Space Astrophysics, Yonsei University, Seoul 120-749, Korea}
\altaffiltext{3}{Department of Physics, University of Oxford, Oxford OX1 3RH, UK}
\altaffiltext{4}{Laboratoire d'Astrophysique de Marseille, 13376 Marseille Cedex 12, France}
\altaffiltext{5}{Department of Physics and Astronomy, UCLA, LA, CA 90095}
\altaffiltext{6}{California Institute of Technology, MC 405-47, Pasadena, CA 91125}
\altaffiltext{7}{Department of Physics and Astronomy, The Johns Hopkins University, Baltimore, MD 21218}
\altaffiltext{8}{Observatories of the Carnegie Institution of Washington, 813 Santa Barbara St., Pasadena, CA 91101}
\altaffiltext{9}{NASA/IPAC Extragalactic Database, California Institute of Technology, Mail Code 100-22, 770 S. Wilson Ave., Pasadena, CA 91125}
\altaffiltext{10}{Department of Astronomy, Columbia University, MC 5246, New York, NY 10027}
\altaffiltext{11}{Department of Astronomy and Space Sciences, Sejong University, Seoul 143-747, Korea}
\altaffiltext{12}{Experimental Astrophysics Group, Space Sciences Laboratory, UC Berkeley, Berkeley, CA 94720}
\altaffiltext{13}{Laboratory for Astronomy and Solar Physics, NASA Goddard Space Flight Center, Greenbelt, MD 20771}

\begin{abstract}

We have used the GALEX UV photometric data to construct a
first near-ultraviolet (NUV) color-magnitude relation (CMR) for the
galaxies pre-classified as early-type by SDSS studies. The NUV CMR is
a powerful tool for tracking the recent star formation history in
early-type galaxies, owing to its high sensitivity to the presence
of young stellar populations. Our NUV CMR for UV-weak galaxies 
shows a well-defined slope and thus will be useful for interpreting
the restframe NUV data of distant galaxies and studying their star
formation history. Compared to optical CMRs, the NUV CMR shows a
substantially larger scatter, which we interpret as evidence of
recent star formation activities. Roughly 15\% of the recent epoch
($z<0.13$) bright ($M[r]<-22$) early-type galaxies show a sign of
recent ($\la$1\,Gyr) star formation at the 1--2\% level (lower 
limit) in mass compared to the total stellar mass. 
This implies that low level residual star formation was common 
during the last few billion years even in bright early-type galaxies. 
\end{abstract}


\section{Introduction}

Color-magnitude relations (CMRs) have been widely-applied tools
for studying the star formation history (SFH) in early-type
galaxies and, in turn, for placing constraints on galaxy 
formation scenarios (monolithic vs. hierarchical).
In optical colors, they show that brighter early-type galaxies 
are generally redder \citep{bau,sv}. 
It is often attributed to a metallicity sequence in the sense that
brighter early-types are more metal-rich \citep{lar,bre,ka}; but
different scenarios, where age plays an important role, have also
been proposed \citep{kc,kav}.

The UV light of an integrated population is a good tracer of
recent star formation (RSF). Thus, finding the local
CMR is important: we can derive the SFH in
early-types by comparing it to the galaxies at various redshifts
\citep{ble,sta,vd99,fs,mae}. The scatter in the CMR is also 
important. It is found to be small in optical CMRs 
\citep{ble}. Because $U$-band light is relatively
sensitive to the presence of young stars, \citet{ble} interpreted
the small scatter as an evidence of absence of recent major star
formation activities in early-types. This result seemed a solid
supporting evidence for the monolithic scenario.

The near-UV (NUV) light is far more sensitive to the presence of 
younger stars than the $U$ band and thus traces RSF history better.
The first internal-release data (IR0.2) from the GALEX project \citep{mar} 
contain the NUV data of nearby galaxies that are large enough 
for statistically significant investigations. It is our goal to 
derive a first NUV vs optical CMR for the present and recent epoch 
early-type galaxies based on this data and investigate their RSF history.

\section{Sample}

GALEX is undertaking two wide-area surveys of different depth \citep{mor}.
The All-sky Imaging Survey (AIS) reaches limiting magnitudes 
19.9AB in the far-UV (FUV, 1344--1786\AA) and 20.8AB in the NUV 
(1771--2831\AA), while the Medium Imaging Survey (MIS) reaches 22.6AB 
and 22.7AB in the FUV and NUV, respectively.  
For nearby galaxies, GALEX operates in the Nearby Galaxy Survey (NGS) 
mode applying the MIS exposure time.
As an illustration, the colors of objects detected in one
MIS field are displayed in Fig.1. Stars (dots mainly in the vertical
sequence in the lower right) are separated from galaxies (crosses) 
clearly. The big galaxy clump in the upper middle of
the plot shows various galaxies currently star-forming, while
quiescent early-type galaxies would be located in the lower part of the
figure in and around the square box. The various symbols show the
expected colors from the Kinney-Calzetti Spectral Atlas of
Galaxies \citep{ck} for redshift 0--0.5 from right to left (arrow). If
early-types are quiescent as often assumed, it would be an easy
task to find them in this two-color diagram.

In order to construct a sample unbiased by any specific search
criterion, we search for GALEX detections of galaxies already
classified as being early-type by one of the major SDSS studies
\citep{ber1}. The Bernardi et al. classification is mainly based 
on concentration index, luminosity profile, and spectra with PCA 
classification.
We use SExtractor's MAG\_AUTO (total) magnitudes from the GALEX catalog
and Bernardi et al.'s {\it model} (total) magnitudes for optical bands.
Our initial cross-identification of sources
between the Bernardi et al. sample and the GALEX catalog results 
in 207 matches.  We have removed 8 and 37 matches for having
a close neighbor within 6 arcsec in GALEX and SDSS images, respectively.
As a result, we have a total of 162 matches (133 from MIS, 18 from AIS,
11 from NGS). Of these, 62 have both FUV and NUV detections. 
Typical errors are 0.1\,mag in NUV and 0.2\,mag in FUV.
These galaxies are at $z = 0$ -- 0.25. It is from this final sample 
that we construct our NUV CMR.

\begin{figure}
\plotone{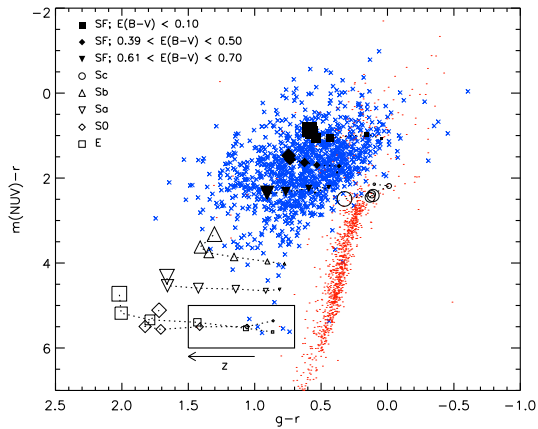} \caption{One GALEX MIS field data show a clear
distinction between stars ($dots$) and galaxies ($crosses$). The
expected positions for typical galaxies from star-forming galaxies
(filled symbols) to quiescent early-type galaxies (open sqaures) are also
depicted for $z = 0$ -- 0.5 (from right to left). Low-redshift
quiescent early-types are supposed to be faint in the UV and
expected to be found in the square box. Optical data are from the
SDSS database. }
\end{figure}

\section{UV CMR}

We use the SDSS $r$ \citep{fuk0} and GALEX NUV photometry to construct
our CMRs (Fig.2). The top panel shows the whole sample, while
the others show galaxies in different redshift bins. The absolute
magnitudes are computed based on the distance derived from redshift, 
assuming $(\Omega, \Lambda, H_{0}) = (0.3, 0.7, 70)$. 
The uncertainty in the redshift is 0.001--0.002, and thus
the uncertainty in the derived distance is negligible \citep{ber1}. 
Extinction is from the Galactic correction of \citet{sch};
we make no correction for internal extinction.
Because we do not assume to know the spectral shapes of these
galaxies a priori, we apply $k$-corrections based on the
luminosity distance only. The $k$-corrections on the colors 
would be 0.1--0.2\,mag. The fitting function to the whole and binned 
samples and the scatter (the standard error between the fit 
and the data), based on the first-order polynomial fitting, are shown 
at the bottom of each panel. 
At first sight, it appears that the slope (dashed line) gets monotonically 
steeper with redshift. But, this is an artifact due to the limiting
magnitude affecting the completeness of the data. The data are
roughly consistent with the global slope in all bins. The change
of the slope is an important issue for studying the SFH
of galaxies, but the sample needs to reach deeper to
assert it. 

\epsscale{1}
\begin{figure}
\plotone{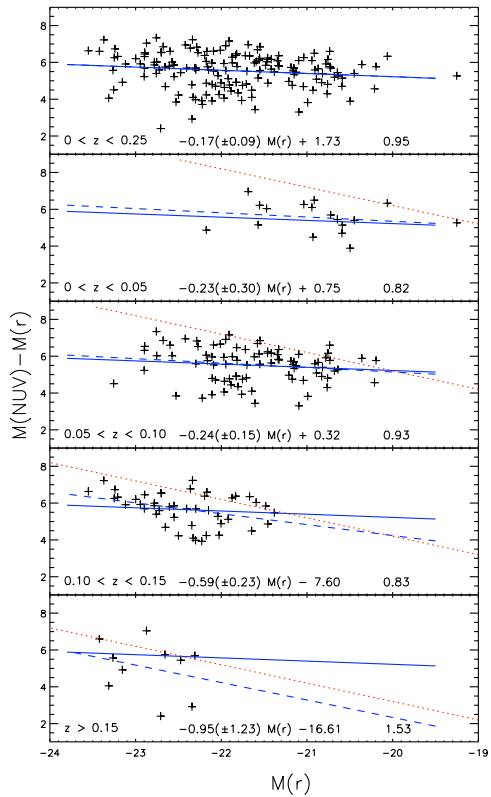} \caption{ The UV CMR for all (top) and various
redshift bins. The slope, intercept, and scatter of the data in
each panel are written at the bottom. The fit to the whole data is
shown in all panels as a solid line; the fit in each redshift bin
is shown as a dashed line; and the detection limit as a dotted line. }
\end{figure}

The most notable feature in the UV CMR is its large scatter. The
scatter in the NUV CMR ($\approx 1$mag) is far greater than that
found in any previous optical CMRs; e.g., the Bernardi et al.
sample shows a $g-r$ scatter $\sigma_{RMS}=0.05$mag \citep{ber4}.
Only part of this scatter can be attributed to the photometric 
uncertainty (0.1mag in NUV) and the $k$-correction on the 
colors (0.15mag at $z=0.2$).
In order to understand the cause of the scatter,
we divide the whole sample into three groups based on the NUV and
FUV fluxes compared to the optical flux. Fig.3 summarizes our
spectral classification scheme. The horizontal lines in the UV for
the continuum fitting (right panels) show the criteria.

\epsscale{1.3}
\begin{figure}
\plotone{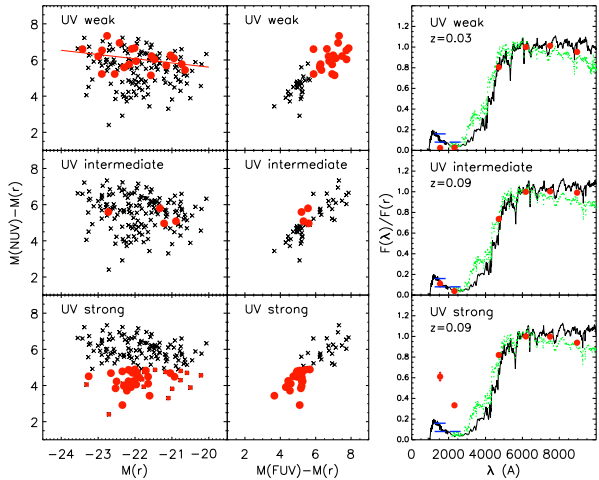} \caption{ Three groups of early-type galaxies
based on the UV spectral shape: ``UV-weak'' (top row), ``UV-intermediate''
(middle), and ``UV-strong'' (bottom). A linear fit to the UV-weak 
galaxies is shown in the top left panel and described in Eqn.1. 
In each panel, $crosses$ are the whole sample, and $circles$ are the 
galaxies in each group, and large $circles$ are those 
with both FUV and NUV detections. In the right panels
(shown in $\lambda$ vs $F_{\lambda}$ format), a typical
example galaxy data (GALEX UV and SDSS optical photometry) for each group
are shown. 
The reference SEDs are those of M32 (dotted) and of NGC\,4552 (solid).
}
\end{figure}

The first group, the ``UV-weak'' galaxies, show a low UV
flux: that is,  $F(NUV)/F(r)<0.08$ and $F(FUV)/F(r)<0.08$.
This originates from the upper bound in the NUV flux measured from 
Burstein et al. (1988)'s nearby quiescent elliptical sample and 
to a flat UV spectrum.
Twenty out of 62 galaxies with both FUV and NUV data 
are classified in this group. 
They form the ``red envelope'' in the UV CMR ($circles$ in Fig.3 top
left panel). The first order polynomial fit to the UV-weak galaxies
is also shown and can be described as
\begin{equation}
M(NUV)-M(r) = -0.23 (\pm 0.16)\, M(r) + 0.95,
\end{equation}
which is consistent with the expected position of the quiescent
ellipticals shown as a box in Fig.1. 
The standard error between the fit and the data is 0.58\,mag.
One can compare this quiescent
early-type galaxy UV CMR to the observed optical CMR of high
redshifts and derive the star formation history. The SED of an
example galaxy is shown in the top right panel, in comparison to
two reference SEDs of NGC4552 (a nearby UV-upturn galaxy) and of M32.
The UV-weak galaxies are commonly suspected to be composed mainly
of old stars.

\epsscale{1.}
\begin{figure}
\plotone{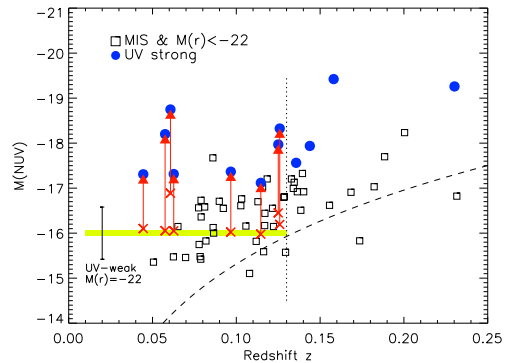} \caption{ Bright ($M[r]<-22$) early-types in
GALEX MIS fields. Filled circles denote the UV-strong galaxies. The
dashed curve marks the detection limit of 23mag. The shaded thick
horizontal line at M(NUV)=$-$16 with error bar ($\sigma$=0.58) 
shows the M(NUV) for quiescent 
galaxies of M(r)=$-$22 expected from the UV-weak fitting function
in Eqn.1. Seven UV-strong early-types are brighter in the UV
than expected by the quiescent models ($crosses$) for their
optical brightness ($M[r]$). Their departure (shown by vertical arrows) 
is probably due to a low level residual star formation in the recent past. }
\end{figure}

We classify 4 out of 62 in the sample with both FUV and NUV detections
as the ``UV-intermediate group''. 
Their NUV flux is low, that is, $F(NUV)/F(r)<0.08$; but their
FUV flux is stronger than NUV, i.e., $0.08<F(FUV)/F(r)<0.16$,
which is typical for nearby elliptical galaxies that show a
UV-upturn \citep{bur}. Extremely old populations may be capable
of developing such a moderate FUV flux \citep{yi,oco};
but, it is difficult to determine whether their UV flux is indeed
caused by young stars or old ones. 
Compared to the FUV light, the NUV light is less affected by the 
presence of old UV-bright stars and makes a safer tracer for RSF.
We simply put these galaxies into 
this category based on the UV spectral criteria without discussing 
their origin in this paper. They occupy the middle part of 
the data below the UV red envelope sequence (middle panels of Fig.3), 
contributing to the scatter of the CMR.

The last ``UV-strong group'' show strong UV flux (bottom right panel) 
either in the NUV $or$ in the
FUV in a manner that is unlikely to have come from old
stars: that is,  $F(NUV)/F(r)>0.08$ $or$ $F(FUV)/F(r)>0.16$. 
Forty four galaxies are classified to be in this group, while
4 of them had no FUV detection.
A sub-group of galaxies show a UV-upturn-type spectral slope, i.e.,
strong in the FUV and weak in the NUV, but a far stronger 
UV flux than any present epoch elliptical galaxy exhibits. 
An example is shown in Fig.3 bottom
right panel. Note that the optical photometric data of this galaxy are 
close to those of typical elliptical galaxies, but its UV flux 
is significantly brighter. Such galaxies
would not appear abnormal in optical CMRs but make themselves
conspicuous in the UV CMR. A 2-component (old and young)
$\chi^2$ test on the GALEX and SDSS data 
indicates a recent starburst age of 0.2\,Gyr
having 1.2\% of the total mass. For 
this test, we assumed that the dominant underlying population
formed at $z=5$ %($t=10.7$\,Gyr at $z=0.12$) 
and of solar metallicity. Their high UV 
flux seems to be hinting at the presence of young (order of 100Myr) stars. 
A larger fraction of galaxies in fact show a strong NUV flux, indicating 
the presence of 0.3 to 1Gyr-old populations. The UV spectral slope 
(FUV$-$NUV) tells us the age of the RSF; and it is clear that the 
UV-strong galaxies are the prime culprit of the scatter in the UV CMR. 
Such a star-formation signature in early-type galaxies has been 
suggested by \citet{deh} who used the FOCA data.

Then, we wonder what fraction of our galaxies are showing RSF signature. 
In order to answer this question, {\it we need to define a subsample
whose red-envelope has been detected}.
Fig.4 shows only the bright ($M[r]<-22$) galaxies from GALEX MIS fields. 
For $z < 0.13$, GALEX reached their red envelope; 
that is, their predicted NUV magnitudes assuming they are UV-weak.
Out of 41 such close bright galaxies, 8 (20\%) are classified as
UV-strong. The $crosses$  
denote their quiescent positions corresponding to their $r$-magnitudes
adopting the UV-weak galaxy CMR in Fig.3. 
Their departure from the quiescent models (vertical arrows)
can be explained by a low level RSF. As a supporting evidence, 
6 of these 8 galaxies show a weak $H_\alpha$ emission line
(see also Salim et al. 2004).

Since galaxy catalogs are in general subject to contamination of 
misclassified objects \citep{gav,shi}, we visually inspected the SDSS 
$r$-band images of all our galaxies. We find most of our 
UV-weak and UV-intermediate
galaxies have early-type morphology but roughly 30\% of UV-strong galaxies
show ambiguous morphology, while still fitting the de 
Vaucouleurs profile. More distant galaxies are obviously more difficult 
to classify. For example, two of the 8 UV-strong galaxies in Fig.4 
appear to be spiral. 
Some galaxies, which clearly appear to be early-type, show minor
non-smooth features. Such features warn us about galaxy misclassification
but might also be expected if the early-type galaxies experienced a 
merger event which caused RSF. 
Further investigation on their morphology, especially using multiband
data, seems essential. 
When the two sprial-looking galaxies are removed from the UV-strong 
sample, our UV-strong fraction becomes 15\% (6/39).

The NUV spectral signature as a sign of RSF remains
apparent only for 1.5\,Gyr or so. In other words, our NUV flux
analysis detects only roughly a 1.5\,Gyr or younger starburst. 
The typical mass fraction of the young populations in these
RSF galaxies is 1--2\%, based on our simple 2-starburst analysis.  
If the recent burst was not instant but extended, the 1--2\% 
estimate should increase. In this regard, this value is a lower limit.

\section{Discussion}

We have constructed a first UV CMR at present and recent epochs at
$z=0$ -- 0.25. One can now compare this empirical UV CMR to higher 
redshift early-type galaxy data and derive their star formation history.
Based on our sample, roughly 15\% of bright ($M[r]<-22$) early-type 
galaxies show a RSF signature in the UV continuum. 
This effectively rules out extreme 
versions of monolithic galaxy formation models where $all$ stars form 
at high redshifts. 
The 15\% estimate is a lower limit in the sense that our sample relies 
on Bernardi et al's 
classification, which excluded galaxies showing strong emission-lines
\citep{fuk}. In addition, the internal extinction, which we ignored
in this study, must have reduced the UV flux in some galaxies.

Recent semi-analytic models based on $\Lambda$CMD dynamics appear
to be compatible with our discovery. For instance, \citet{kav} 
suggested that roughly 5--10\% --- depending strongly on the galaxy mass 
and environment --- of the entire star formation in all (bright and 
faint) early-types happened at $z$$<$1. The meger rate and the
amount of residual star formation are predicted to be a sharply-increasing 
function of redshift \citep{kb}, which is supported by observational 
studies \citep{mae,bel}. A detailed test against galaxy evolution 
models using GALEX data will be presented shortly. 

An obvious next question is what triggers residual star formation. 
If it is a merger event, some UV-strong galaxies with a very young 
RSF might show morphologically disturbed features, as shown in HI gas 
map of the nearby elliptical galaxy, Cen~A \citep{sch94}. A follow-up
investigation would be called for.

The strength of the UV CMR study is that it can detect even sub-1\% level
RSF event very easily. During its 28-month mission, GALEX will collect
an order of magnitude larger and deeper data than used here and provide 
critical information on the recent star formation history in early-type 
galaxies.

\section*{Acknowledgments}

GALEX (Galaxy Evolution Explorer) is a NASA Small Explorer, launched 
in April 2003. We gratefully acknowledge NASA's support for construction, 
operation, and science analysis for the GALEX mission, developed in 
cooperation with the Centre National d'Etudes Spatiales
of France and the Korean Ministry of Science and Technology.
We thank Mariangela Bernardi for useful information on her catalog
and Sadegh Khochfar, Joseph Silk, Roger Davies, Ignacio Ferreras 
for valuable comments. We thank the referee for a number of constructive
suggestions.

\end{document}